\documentclass[a4paper]{article}

\usepackage{icrc2013}
 \usepackage[english]{babel}
 
\title{The Cherenkov Telescope Array Large Size Telescope}

\shorttitle{ICRC 2013 Template}

\authors{
G.~Ambrosi$^{17}$,
Y.~Awane$^{12}$,
H.~Baba$^{6}$,
A.~Bamba$^{1}$,
M.~Barcel\'{o}$^{8}$,
U.~Barres~de~Almeida$^{3}$,
J.A.~Barrio$^{25}$,
O.~Blanch~Bigas$^{8}$,
J.~Boix$^{8}$,
L.~Brunetti$^{13}$
% ...
for the Cherenkov Telescope Array Consortium.
}

\afiliations{
$^1$ Department of Physics and Mathematics, Aoyama Gakuin University, Fuchinobe, Sagamihara, Kanagawa, 229-8558, Japan\\
% ...
$^3$ Centro Brasileiro de Pesquisas Fisicas (CBPF/MCTI), Rua Xavier Sigaud 150, RJ 22290-180, Rio de Janeiro, Brazil\\
% ...
$^6$ Faculty of Science, Ibaraki University, Mito, Ibaraki, 310-8512, Japan\\
% ...
$^8$ Institut de F\'{i}sica d'Altes Energies (IFAE), Edifici Cn, Campus UAB, 08193 Bellaterra, Spain\\
% ...
$^{12}$ Kyoto University, Sakyo-ku, Kyoto 606-8502, Japan\\
$^{13}$ LAPP, Universit\'{e} de Savoie, CNRS/IN2P3, 9 Chemin de Bellevue - BP 110, 74941 Annecy-le-Vieux Cedex, France\\
% ...
$^{17}$  INFN Sezione di Perugia, Via A. Pascoli, Perugia, Italy\\
% ...
$^{25}$ Grupo de Altas Energ\'{i}as, Universidad Complutense de Madrid, Av Complutense s/n, 28040 Madrid, Spain\\
% ...
}

\email{blanch@ifae.es}

\abstract{The two arrays of the Very High Energy gamma-ray observatory Cherenkov
Telescope Array (CTA) will include four Large Size Telescopes (LSTs) each with a 23 m diameter dish and 28 m focal distance. These telescopes will enable CTA to achieve a low-energy threshold of 20 GeV, which is critical for important studies in astrophysics, astroparticle physics and cosmology. This work presents the key specifications and performance of the current LST design in the light of the CTA scientific objectives.}


\begin{document}
\maketitle

%Begin a section.
\section{CTA Large Size Telescope}

During the past few years, Very High Energy (VHE) gamma ray astronomy has made spectacular progress and has established itself as a vital branch of astrophysics. To advance this field even further, we propose the Cherenkov Telescope Array (CTA) \cite{bib:CTA}, the next generation VHE gamma ray observatory, in the framework of a worldwide, international collaboration. CTA is the ultimate VHE gamma ray observatory, whose sensitivity and broad energy coverage will attain an order of magnitude improvement above those of current Imaging Atmospheric Cherenkov Telescopes. By observing the highest energy photons known, CTA will clarify many aspects of the extreme Universe, including the origin of the highest energy cosmic rays in our Galaxy and beyond, the physics of energetic particle generation in neutron stars and black holes, as well as the star formation history of the Universe. CTA will also address critical issues in fundamental physics, such as the identity of dark matter particles and the nature of quantum gravity.

CTA consists of three types of telescopes to cover a broader energy band, Large Size Telescopes (LSTs,  23m diameter), Mid Size Telescopes (12~m), and Small Size Telescopes (4-6~m). The purpose of LST is to enhance the sensitivity below 200-300GeV and to lower the effective threshold down to 20-30GeV. The science case of LST is the observation of high redshift AGNs up to z $\le$ 3, GRBs up to z $\le$ 10, and pulsars and galactic transients. LST surely expands the domain of science to the cosmological distances and fainter sources with soft energy spectra.

\section{The Structure of the Large Size Telescope}

The telescope geometry is optimized to maximize the cost performance by Monte Carlo simulations and toy models. The baseline parameters are defined with the dish size of 23~m and the focal length of 28m, leading then to f/d = 1.2.

\subsection{Azimuth System}
\begin{figure}[h]
  \centering
  \includegraphics[width=0.4\textwidth]{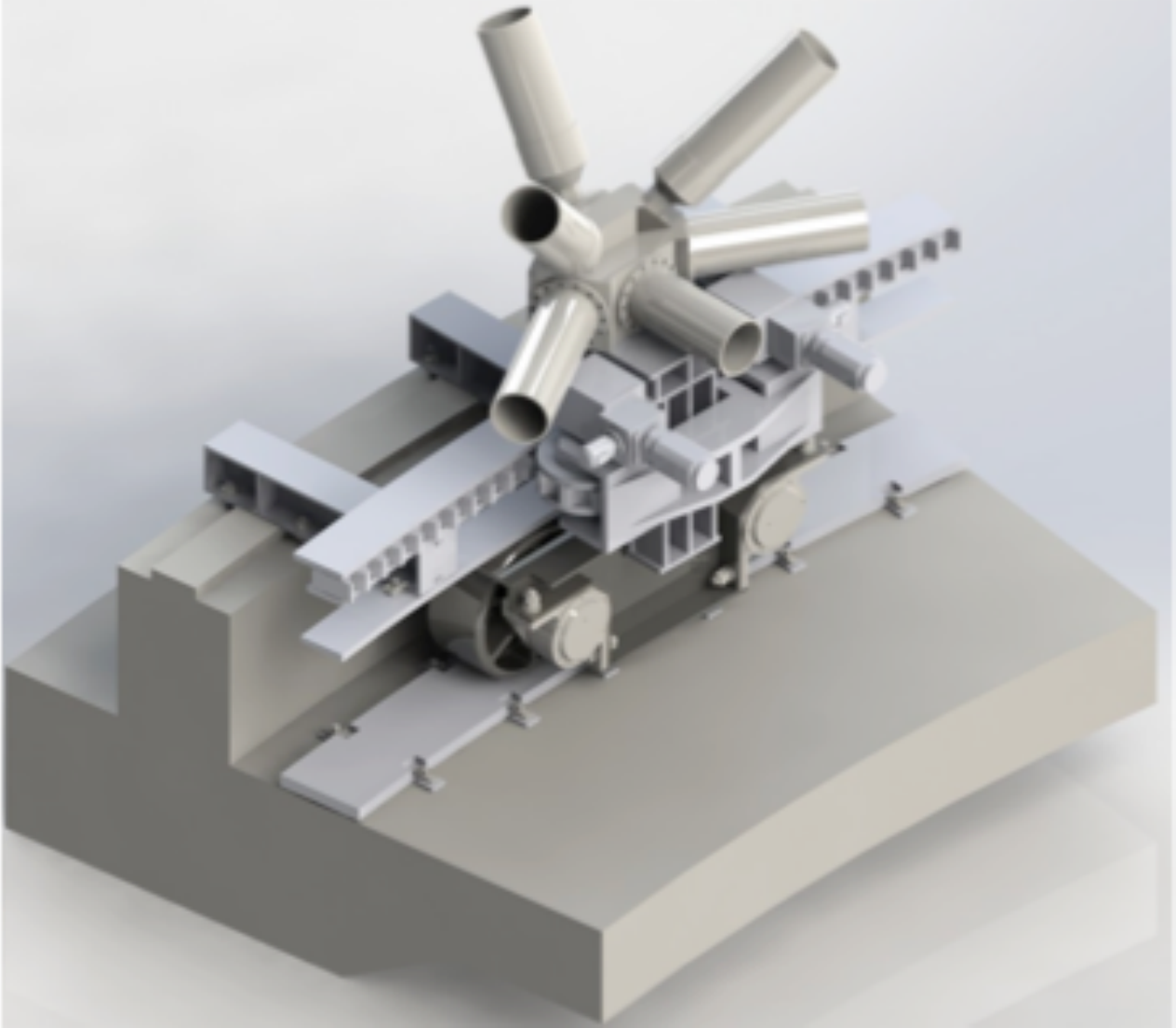}
  \caption{Design detail of the azimuth system.}
  \label{bogie_fig}
 \end{figure}

The main objective of LST azimuth system is to allow the telescope to turn along its vertical axis. It has been designed by the CTA-Spain consortium. The telescope structure rests on six bogies equally spaced in a hexagonal array.  Each bogie has two wheels that turn in a double rail system (figure \ref{bogie_fig}). The azimuth driving system is an inner spur gear (crown) of about 24 m diameter. The two bogies withstanding most of the weight have two pinions powered by servomotors, turning the telescope. The spur gear assures that the azimuth system turns always, either if the bogie wheels are sliding or not. Bogies and rail are covered in order to extend the azimuth system lifespan.

\subsection{The azimuth substructure of the telescope}
\begin{figure}[h]
  \centering
  \includegraphics[width=0.4\textwidth]{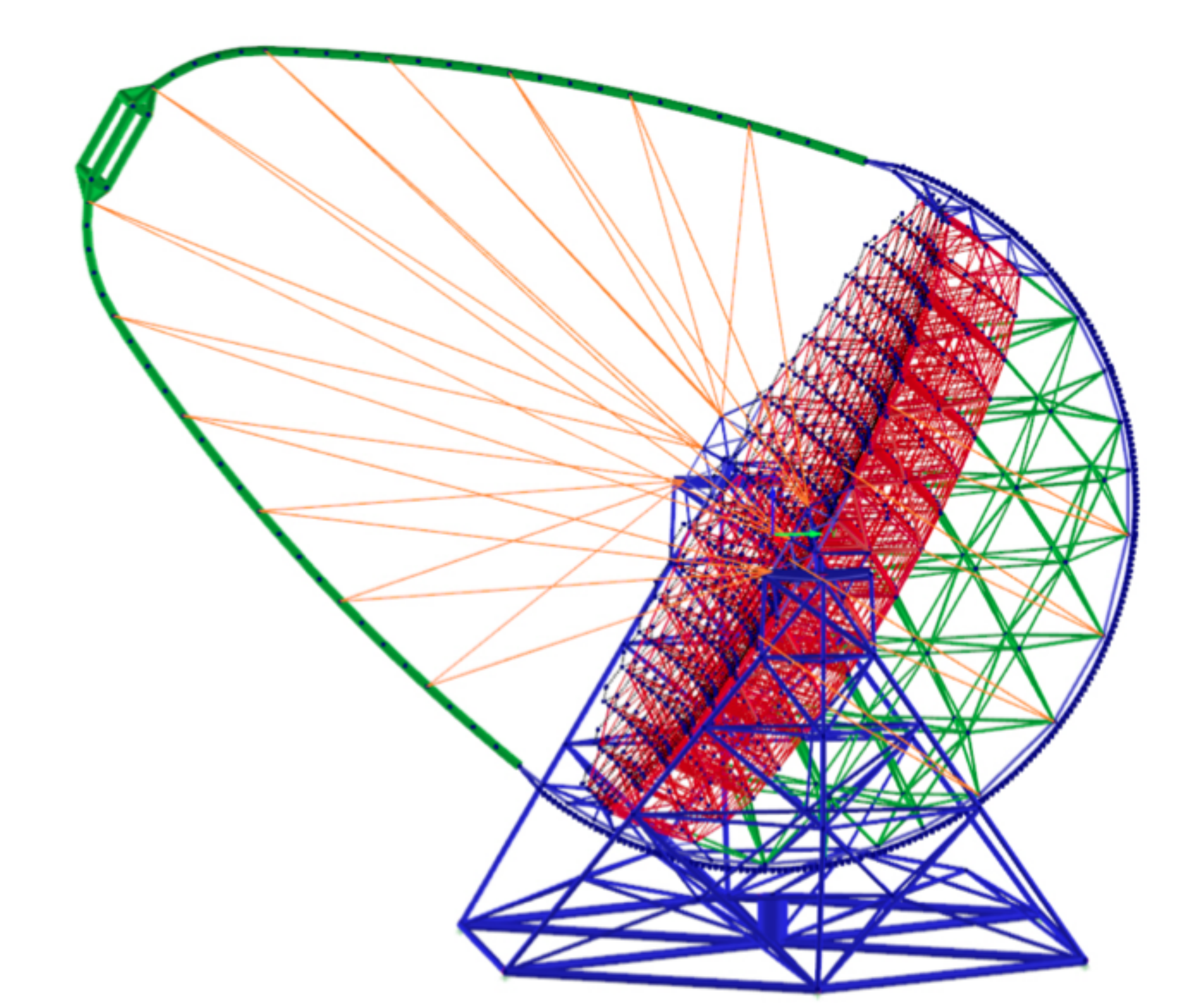}
  \caption{Structure of the telescope mount (simplified, without bogies, mirrors and camera).}
  \label{structure_fig}
 \end{figure}

The azimuth substructure (lower part of figure \ref{structure_fig}) is a space frame structure with carbon fiber reinforced plastic (CFRP) tubes designed by the MPI Munich group together with MERO-TSK. It is basically a copy of the azimuth substructure of the MAGIC telescope\cite{bib:MAGIC}. The main difference is the use of large diameter CF tubes instead of steel tubes to lower considerably the weight of the telescope down to 60-65 tons. The design comprises 134 tubes of which only eight are reinforced steel tubes. These tubes connect the bogies with each other and the central axis.

\subsection{The mirror support dish}

The mirror support dish is a double layer space frame and also follows closely the concept of the MAGIC telescopes\cite{bib:MAGIC}. The basic LST elements are CFRP tubes of either 80 or 100 mm diameter arranged in a tetrahedral structure using the patented MERO construction principle. The proposed tetrahedral structure is the stiffest basic element. Moreover, the tetrahedral configuration offers an ideal support for the planned hexagonal mirror panels, which can be fixed at the three corners of the tetrahedra of the top layer close to the nodes. The nodes are made from aluminum spheres and each one machined with threads such that a nearly parabolic profile can be approximated. In total the space frame will comprise 2630 CF tubes and 950 nodes. The breaking strength of one tube is expected to be over 80 tons at 35$^{\circ}$C, hence a single tube should be strong enough to hold the entire telescope.

\subsection{Elevation system}
Across the backside of the dish the space frame structure is extended to carry the semicircular drive ring.  This space frame section is made in part by heavier steel tubes to help counterweight the telescope support structure. The declination drive ring is formed by a bent I-beam acting as a support for a chain bolted ever 5~cm to the beam. The I-beam has a machined surface to act as a rail for the box carrying the 10~kW declination drive motor and a disc brake to prevent the dish rotation in case of a motor failure.

\subsection{Camera Support Structure}

The Camera Support Structure \cite{bib:trigger}, conceived and designed by the LAPP group, responds to well defined optical specifications of the telescope by minimizing the weight of the masts and the shadowing of the primary mirror. The arch design consists on three curved sections for each of the arms using Carbon Fibre Plastic Reinforced, with a 310 mm diameter circular cross section and a constant thickness of 14~mm. There are 26 ropes used to stiffen the structure. The cables fixation devices on the arch are well-known by naval designers and very commonly used to link masts to cables. The camera frame has an internal square space of 3140x3140mm, allocated for the sliding and fixing of the octagonal camera, and the square lids.

 %\begin{figure}[h]
  %\centering
  %\includegraphics[width=0.4\textwidth]{icrc2013-0776-03}
 % \caption{Modelization of the camera support structure design.}
 % \label{CSS_fig}
% \end{figure}

\subsection{Drive system and slow control architecture}
The requirements of fast and precise positioning of the LSTs will be fulfilled by driving the elevation and azimuth axis of each single telescope by electric servo motors. Due to heavy torque and the precision required, synchronous motors type will be used on the LST telescope. A maximum total mechanical power of 190~Kw (145~Kw for the azimuth drive and 45~Kw for the elevation drive) is needed. For the Azimuth Axis, since we use 4 synchronous motors with 4 times overload capabilities, 12~Kw individual motors are sufficient to drive the LST.

The Slow Control architecture proposed is based on a set of devices (PLCs, industrial PCs and smart embedded devices (sensors/actuators). They will be able to provide homogeneous command and control data access through a standardized software layer for all telescope types. A generic description of each telescope system and subsystem behavior will enable a homogeneous Array Control for Slow Control dataflow.

\section{Reflector system}

\subsection{The Mirrors}
The global mirror structure of the LST should be a parabolic shape to keep the isochronousity of the optics. The entire reflector with its diameter of 23~m consists of 198 hexagonal mirrors with a flat to flat dimension of 1.5~m. Individual mirrors have a spherical shape, but we may have two or three groups with slightly different focal lengths. Shorter focal length mirrors at the central area of the reflector and longer ones at outer area can be arranged. The total area of the reflector is about 396~m$^{2}$.

\begin{figure}[h]
  \centering
  \includegraphics[width=0.4\textwidth]{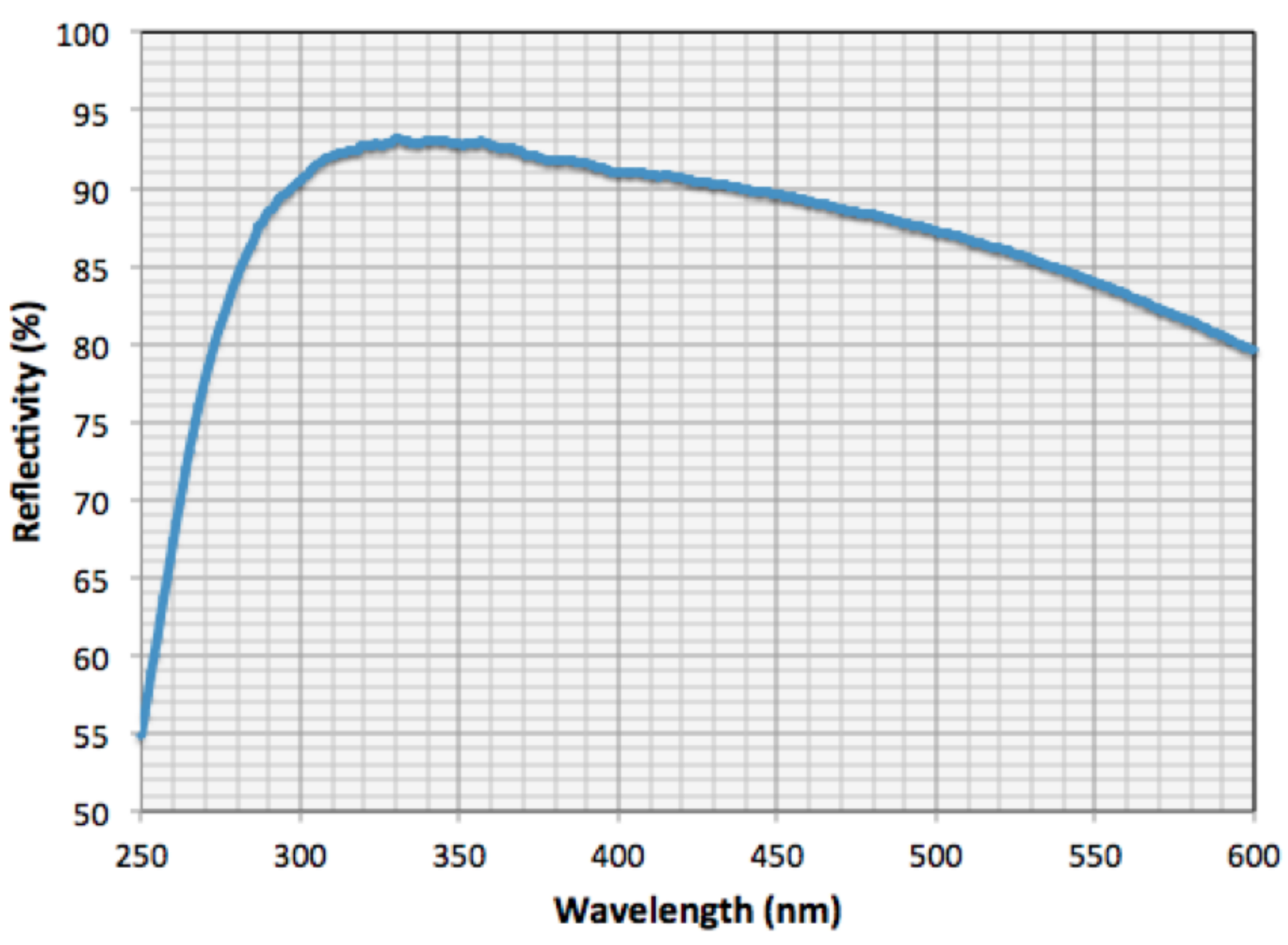}
  \caption{The reflectivity of a mirror with five-layer coat (Cr, Al, SiO2, HfO2, SiO2) produced at Sanko.}
  \label{MirrorReflectivity_fig}
 \end{figure}

The segmented mirrors will be fabricated by the Japanese company Sanko with the cold slump technique. They have a sandwich structure consisting of glass sheet of 2.7~mm thickness Ð aluminum honeycomb of 60~mm thickness - glass sheet. The reflective layer of the mirror is coated with Cr and Al on the surface of the glass sheet with a protective multi-coat layer of SiO2, HfO2 and SiO2. These five layers are produced with the spattering technology. By adjusting the thickness of individual layers with SiO2 and HfO2, we can optimize the reflectivity to 90\% or higher at 400~nm (figure \ref{MirrorReflectivity_fig}).

\subsection{Active mirror control system}

The telescope suffers deformations and bending due to several causes. All dish deformations, including a change of the focal distance, can be corrected by the active mirror control system. Each hexagonal mirror panel for LST will be fixed on three neighboring knots. Two of the points are mounted on actuators and one point is fixed. On each knot of the telescope dish a mounting plate is fixed such that each mirror can be fixed on three points. This construction allows an adjustment of each mirror panel in two directions. The automatic adjustment needs a feedback loop. The optical axis of the LST will be defined with two infra-red lasers at the center of the dish constantly shining two targets left and right of the imaging camera. Infrared lasers will also be mounted on the edge of each mirror segment. The directional offset of the mirror facets will be estimated by taking pictures of the spots on the target near the Camera with a high resolution IR CCD camera viewing from the center of the dish. The actuator consists of a stepper motor attached to a spindle and will move the segmented mirrors to correct the deformations. The achievable positioning resolution is $<$ 5~$\mu$m.

\section{Imaging Camera with PMTs}

Based on simualtion studies, the LST camera will have a field of view of 4.5$^{\circ}$. It can conceptually be divided in three different parts: the Focal Plane Instrumentation, the cluster electronics and the global camera elements. All three parts go inside a sealed structure with temperature control.
 
%In total, we divide the camera in five almost independent blocks: Front, FPI, Clusters, Global camera elements and Back.

%\subsection{The camera front}

\subsection{Camera mechanics and cooling system}

\begin{figure}[t]
  \centering
  \includegraphics[width=0.4\textwidth]{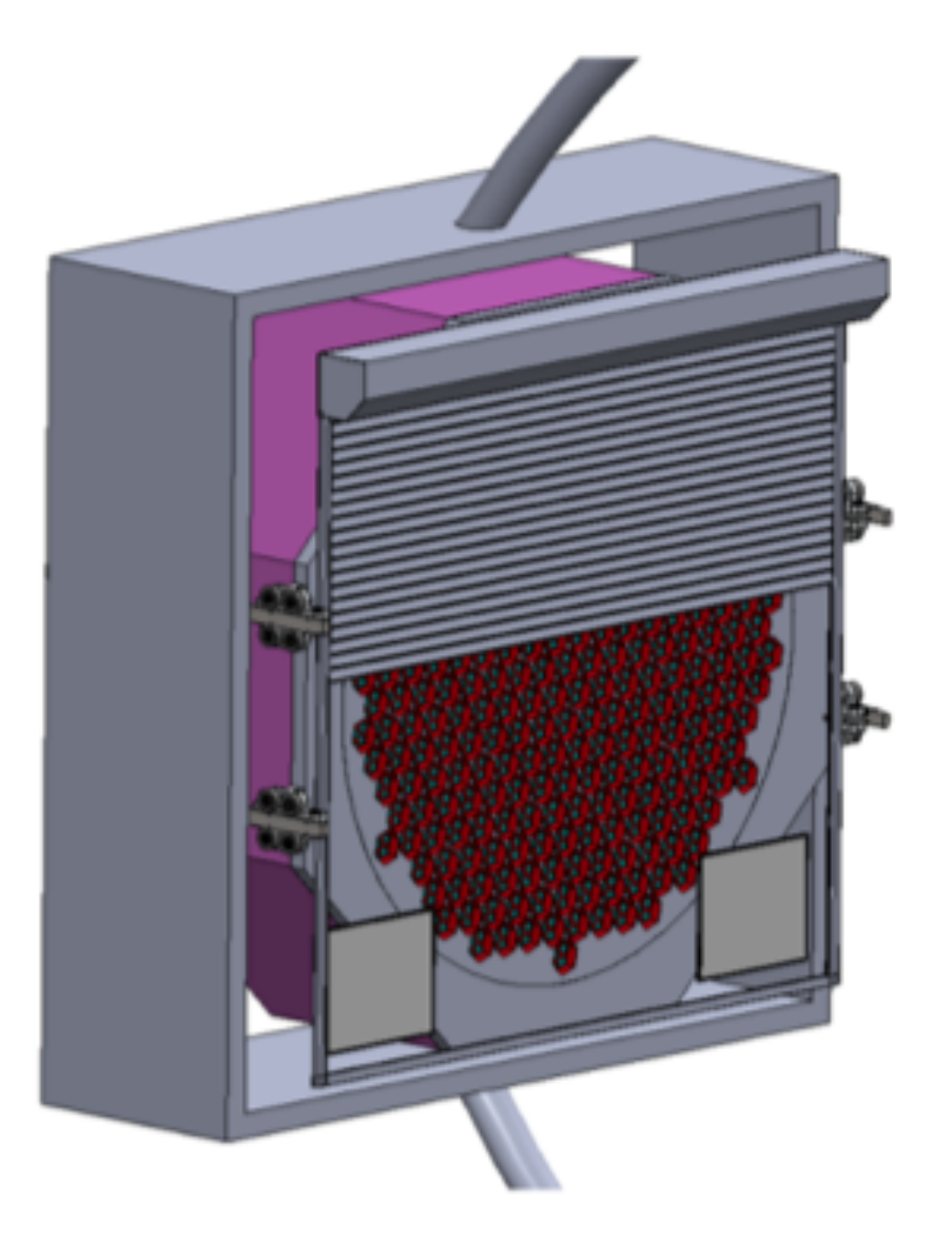}
  \caption{External view of LST camera design,}
  \label{Mechanics_fig}
 \end{figure}

The camera mechanical structure \cite{bib:CamMech} requirements are defined by the environmental working conditions, the specifications on the required environmental conditions inside the camera, and on the positioning requirements of the camera pixels. Moreover, a strong constraint comes from the structure of the LST which requires that the total camera weight not exceed 2000~kg. It is attached to the mast frame structure using a rails system, which allows to modify the camera position with respect to the mirror, and to unload it for maintenance. In order to protect the photomultipliers and allow operations during daylight with high voltage switched on, a motorized shutter completely closes the entrance of light from the front of the camera when required (see figure \ref{Mechanics_fig}).

The power dissipation of the electronics inside the camera can reach 5kW, which results in a considerable amount of heat that has to be taken out from the camera body in order to keep all elements within their working range in stable temperature conditions. Two different cooling systems have been considered and are being studied: a temperature controlled air flow cooling system, and a water cooling system based on cold plates\cite{bib:CamMech}.

\subsection{Focal Plane Instrumentation}

A photodetector module consists of a light guide, photomultiplier, Cockcroft Walton high voltage power supply and preamplifier. These components make a module and work as a pixel on the focal plane. 

The photo-detector modules are mounted with 50~mm spacing, while the sensitive area of the photomultiplier cathode is a circular area with a diameter of 30~mm. Therefore a light guide is essential to collect all Cherenkov light focused on the camera plane. The entrance window of the light guide is a regular hexagon with the size of 50~mm in flat to flat. The shape of the light guide and the reflective material, 3M ESR film, are optimized to get the maximum efficiency up to 26.7 degrees, the maximum incidence angle given the F/D of the mirror and Þeld of view of the camera.

Hamamatsu R11920-100-20 is the photomultiplier developed for the CTA project. This PMT has 8 Dynodes, a 40 mm diameter window, and a Super Bialkali Cathode with a radius of curvature of 21 mm. Frosted entrance glass is used for the PMT surface window. These photomultipliers were developed aiming at large Quantum Efficiency and low after pulse rates. The former reaches values of 47\% and the latter goes below 0.02 \% above 4 phe. The Cockcroft-Walton is also provided by Hamamatsu and can supply high voltage from 850~V to 1500~V.

Finally, the signal goes through PACTA, developed by CTA-Spain, before reaching the cluster electronics. PACTA is a wide dynamic range preamplifier with low power consumption and low noise. 

In order to fulfill the accessibility and weight requirements, the camera body relies on a simple structural element, or load bearing structure, which provides the necessary stiffness and positioning accuracy for the photo-detectors, and holds all the camera elements.

\subsection{Cluster electronics}

\begin{figure}[t]
  \centering
  \includegraphics[width=0.4\textwidth]{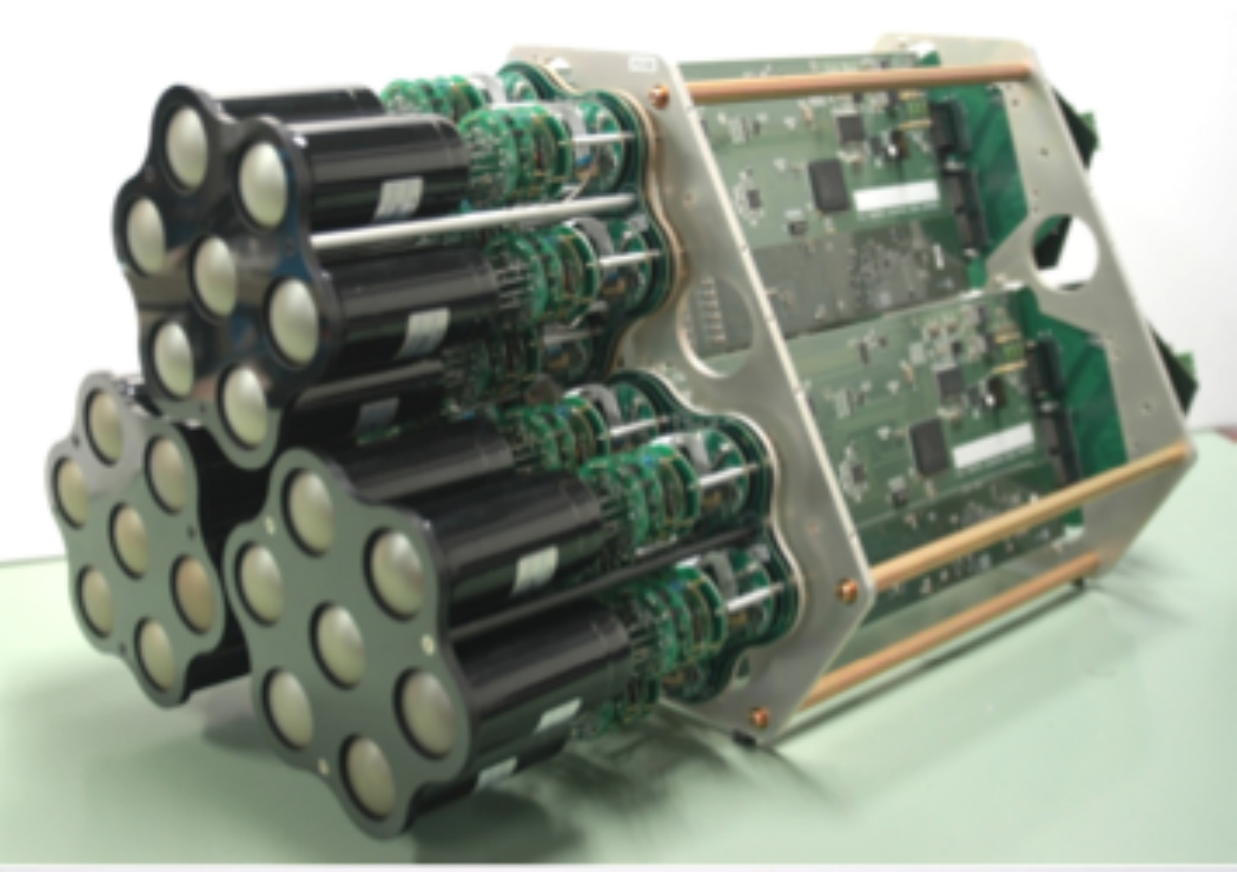}
  \caption{Three clusters including the photodetector modules and electronics.}
  \label{Dragon_fig}
 \end{figure}

The photodetector modules are arranged on the focal plane of each telescope, with one readout system per 7-modules cluster \cite{bib:dragon} (figure \ref{Dragon_fig}). The readout system developed by CTA-Japan is based on the analogue memory DRS version 4 (DRS4) \cite{bib:Ritt}. On the readout board the signal is divided into three lines: a high gain channel, a low gain channel, and a trigger channel. The high and low gain channels are connected to DRS4 chips. The signal is sampled at a rate of the order of GHz and the waveform is stored in the analogue memory. When a trigger is generated, the voltages stored in the capacitor array are sequentially output and then digitized by an external slow sampling (~30 MHz) ADC. 

Although several hardware developments are being developed and will be tested with several clusters, the simulations show a superior performance of the Sum Trigger concept \cite{bib:Rissi} at the lowest energies, which are the most critical for the LST. Hence, the implementation of the Sum Trigger concept  done by CTA-Spain is the baseline option\cite{bib:trigger}.

\vspace*{0.5cm}
\footnotesize{{\bf Acknowledgment:}{We gratefully acknowledge financial support from the agencies and organisations listed in this page: http://www.cta-observatory.org/?q=node/22.}}

\end{document}